\def\Circe/{\texttt{circe}}
\def\Version/{1.0}
\def\Date/{July 1996}
\documentclass[a4paper]{article}
\usepackage{amsmath}
\allowdisplaybreaks
\usepackage{graphics}
\usepackage{noweb}
\def\nwendcode{\endtrivlist\endgroup}
\let\nwdocspar\relax
\IfFileExists{thohacks.sty}%
  {\usepackage{thohacks}}%
  {}%
\IfFileExists{thogreek.sty}%
  {\usepackage{greek}%
   \def\Kirke/{{\greek Ki'rkh}}%
   \def\Hydra/{{\greek ('Udra}}%
   \def\Ulysses{%
     \begin{flushright} % Odyssee: 10,320
       \greek e)'rxeo nu=n sufeo'nde, met'' a)'llwn le'co e(tai'ron.
     \end{flushright}%
     \begin{flushright} % Ulysses: 15,2844-2845
       \textsc{Bloom:} \textit{(with sinews semiflexed)} Magmagnificence!
     \end{flushright}}}%
  {\def\Kirke/{K$\acute\iota${}$\rho${}$\kappa${}$\eta$}%
   \def\Hydra/{Y$\delta${}$\rho${}$\alpha$}%
   \let\Ulysses\relax}%
\IfFileExists{mflogo.sty}%
  {\usepackage{mflogo}}%
  {\def\MF{\textsf{META}\-\textsf{FONT}}}%
 {\begin{list}{}%
   {\setlength{\leftmargin}{2em}%
    \setlength{\rightmargin}{2em}%
    \setlength{\itemindent}{-1em}%
    \setlength{\listparindent}{0pt}%
    }}%
 {\end{list}}
\makeindex
\begin{document}
\bibliographystyle{prsty}%%%{unsrt}%%%{physics}
\title{%
  \Kirke/ Version \Version/:\\
  Beam Spectra for Simulating Linear Collider Physics%
    \thanks{%
       Supported by Bundesministerium f\"ur Bildung,
       Wissenschaft, Forschung und Technologie, Germany.}}
\author{%
  Thorsten Ohl%
    \thanks{e-mail: \texttt{Thorsten.Ohl@Physik.TH-Darmstadt.de}}\\
  \hfil\\
    Technical University of Darmstadt\\
    Schlo\ss{}gartenstr. 9\\
    D-64289 Darmstadt\\
    Germany}
\date{%
  IKDA 96/13\\
  hep-ph/9607454\\
  July 1996}
\maketitle
\begin{abstract}
  I describe parameterizations of realistic $e^\pm$- and $\gamma$-beam
  spectra at future linear $e^+e^-$-colliders.  Emphasis is put on
  simplicity and reproducibility of the parameterizations, supporting
  reproducible physics simulations. The parameterizations are
  implemented in a library of distribution functions and event
  generators.
\end{abstract}
%%%%%%%%%%%%%%%%%%%%%%%%%%%%%%%%%%%%%%%%%%%%%%%%%%%%%%%%%%%%%%%%%%%%%%%%
\newpage
\section*{Program Summary:}
\begin{itemize}
  \item \textbf{Title of program:}
    \Kirke/, Version \Version/ (\Date/)
%CPC%  \item \textbf{Catalogue number:}
%CPC%    ????
  \item \textbf{Program obtainable}
%CPC%    from CPC Program Library, Queen's University of Belfast,
%CPC%    N.~Ireland (see application form in this issue) or
    by anonymous \verb|ftp| from the host
    \hfil\allowbreak\verb|crunch.ikp.physik.th-darmstadt.de|
    in the directory
    \allowbreak\verb|pub/ohl/circe|.
  \item \textbf{Licensing provisions:}
    Free software under the GNU General Public License.
  \item \textbf{Programming language used:}
    Fortran77
  \item \textbf{Number of program lines in distributed program, including
      test data, etc.:}
    $\approx$ 1100 (excluding comments)
  \item \textbf{Computer/Operating System:}
    Any with a Fortran77 programming environment.
  \item \textbf{Memory required to execute with typical data:}
    Negligible on the scale of typical applications calling the library.
  \item \textbf{Typical running time:}
    A small fraction (typically a few percent) of the running time of
    applications calling the library.
  \item \textbf{Purpose of program:}
    Provide simple and reproducible, yet realistic, parameterizations
    of the $e^\pm$- and $\gamma$-beam spectra for linear colliders.
  \item \textbf{Nature of physical problem:}
    The intricate beam dynamics in the interaction region of a
    high luminosity linear collider at $\sqrt s = 500$GeV result in
    non-trivial energy spectra of the scattering electrons, positrons and
    photons.  Physics simulations require simple and reproducible, yet
    realistic, parameterizations of these spectra.
  \item \textbf{Method of solution:}
    Parameterization, curve fitting, Monte Carlo event generation. 
  \item \textbf{Keywords:}
    Event generation, beamstrahlung, linear colliders.
\end{itemize}
\newpage
\nwfilename{}\nwbegindocs{1}%
%%%%%%%%%%%%%%%%%%%%%%%%%%%%%%%%%%%%%%%%%%%%%%%%%%%%%%%%%%%%%%%%%%%%%%
\Ulysses
\section{Introduction}
Despite the enormous quantitative success of the electro-weak standard
model up to energies of $200\text{GeV}$, neither the nature of
electro-weak symmetry breaking~(EWSB) nor the origin of mass are
understood.\par
{}From theoretical considerations, we know that clues to the answer
of these open questions are hidden in the energy range
below~$\Lambda_{\text{EWSB}} = 4\pi v \approx 3.1\text{TeV}$.  Either
we will discover a Higgs particle in this energy range or signatures
for a strongly interacting EWSB sector will be found.  Experiments at
CERN's Large Hadron Collider~(LHC) will shed a first light on this
regime in the next decade.  In the past is has been very fruitful to
complement experiments at high energy hadron colliders with
experiments at $e^+e^-$-colliders. The simpler initial state allows
more precise measurements with smaller theoretical errors.  Lucid
expositions of the physics opportunities of high energy $e^+e^-$
colliders with references to the literature can be found
in~\cite{Murayama/Peskin:1996:LC_review}.\par
However, the power emitted by circular storage rings in form of
synchrotron radiation scales like~$(E/m)^4/R^2$ with the energy and
mass of the particle and the radius of the ring.  This cost becomes
prohibitive after LEP2 and a Linear Collider~(LC) has to be built
instead.\par
Unfortunately, the ``interesting'' hard cross sections scale
like~$1/s$ with the square of the center of mass energy and a LC
will have to operate at extremely high luminosities in excess
of~$10^{33}\text{cm}^{-2}\text{s}^{-1}$.  To achieve such luminosities,
the bunches of electrons and positrons have to be very dense. Under
these conditions, the electrons undergo acceleration from strong
electromagnetic forces from the positron bunch (and vice versa).  The
resulting synchrotron radiation is called
\emph{beamstrahlung}~\cite{Chen/Noble:1986:Beamstrahlung}
and has a strong effect on
the energy spectrum~$D(x_1,x_2)$ of the colliding particles.  This
changes the observable $e^+e^-$~cross sections
\begin{subequations}
\begin{align}
  \frac{d\sigma^{e^+e^-}_0}{d\Omega}(s) \to
  \frac{d\sigma^{e^+e^-}}{d\Omega}(s) &=
    \int_0^1 \!dx_1dx_2\, D_{e^+e^-} (x_1,x_2;\sqrt s)
       J(\Omega',\Omega) \frac{d\sigma^{e^+e^-}_0}{d\Omega'}(x_1x_2s)\\
\intertext{%
  and produces luminosity for $e^\pm\gamma$ and
  $\gamma\gamma$~collisions:}
  \frac{d\sigma^{e^\pm\gamma}}{d\Omega}(s) &=
    \int_0^1 \!dx_1dx_2\, D_{e^\pm\gamma} (x_1,x_2;\sqrt s)
       J(\Omega',\Omega) \frac{d\sigma^{e^\pm\gamma}_0}{d\Omega'}(x_1x_2s)\\
  \frac{d\sigma^{\gamma\gamma}}{d\Omega}(s) &=
    \int_0^1 \!dx_1dx_2\, D_{\gamma\gamma} (x_1,x_2;\sqrt s)
       J(\Omega',\Omega) \frac{d\sigma^{\gamma\gamma}_0}{d\Omega'}(x_1x_2s)
\end{align}
\end{subequations}
Therefore, simulations of the physics expected at a LC
need to know the spectra of the~$e^\pm$ and~$\gamma$ beams
precisely.\par
Microscopic simulations of the beam dynamics are available
(e.g.~\texttt{ABEL}\cite{Yokoya:1985:ABEL},
\texttt{CAIN}\cite{Chen/etal:1995:CAIN}
and~\texttt{Guinea-Pig}\cite{Schulte:1996:Thesis}) and their
predictions are compatible with each other.
But they require too much computer time and memory for direct use in 
physics programs.  \Kirke/ provides a fast and simple parameterization
of the results from these simulations.  Furthermore, even if the
computational cost of the simulations would be negligible, the input
parameters for microscopic simulations are not convenient for particle
physics applications.  Due to the highly non-linear beam dynamics,
the optimization of LC designs is a subtle
art~\cite{Palmer:1990:LC_review}, that is best practiced by the
experts.  Furthermore, particle physics applications
need benchmarking and \text{easily reproducible}
parameterizations are required for this purpose.\par
The parameterizations in \Kirke/ are not based on approximate
solutions~(cf.~\cite{Chen:1992:Beamstrahlung}) of the beamstrahlung
dynamics.  Instead, they provide a ``phenomenological'' description of
the results from full simulations.  The parameterizations are as simple
as  possible while remaining consistent with basic physical principles:
\begin{enumerate}
  \item \emph{positivity:} the distribution functions~$D(x_1,x_2)$
    \emph{must not} be negative in the physical
    region~$[0,1]\times[0,1]$.
  \item \emph{integrability:} the definite integral of the
    distribution functions over the physical
    region~$[0,1]\times[0,1]$ \emph{must} exist, even though the
    distributions can have singularities.
\end{enumerate}
This paper is organized as follows: I start in
section~\ref{sec:parameters} with a discussion of the input for the
microscopic simulations.  In section~\ref{sec:usage} I describe the
usage of the \Kirke/ library and in
section~\ref{sec:technical} I discuss some technical details of the
implementation.   After discussing the parameterizations available in
version \Version/ in section~\ref{sec:parameterizations}, I conclude
in section~\ref{sec:conclusions}.
\nwenddocs{}%
\nwbegindocs{2}%
%%%%%%%%%%%%%%%%%%%%%%%%%%%%%%%%%%%%%%%%%%%%%%%%%%%%%%%%%%%%%%%%%%%%%%
\section{Parameters}
\label{sec:parameters}
%%%%%%%%%%%%%%%%%%%%%%%%%%%%%%%%%%%%%%%%%%%%%%%%%%%%%%%%%%%%%%%%%%%%%%%%
\begin{table}
  \begin{center}
    \renewcommand{\arraystretch}{1.3}
    \begin{tabular}{|c||c|c|c|c|c|c|}\hline
          & \texttt{SBAND} & \texttt{TESLA} & \texttt{XBAND}
          & \texttt{SBAND} & \texttt{TESLA} & \texttt{XBAND}
      \\\hline\hline
      $E/\text{GeV}$                  & 250    &  250    & 250    
                                      & 500    &  500    & 500    \\\hline
      $N_{\text{particles}}/10^{10}$  &   1.1  &    3.63 &   0.65 
                                      &   2.9  &    1.8  &   0.95 \\\hline
      $\epsilon_x/10^{-6}\text{mrad}$ &   5    &   14    &   5    
                                      &  10    &   14    &   5    \\\hline
      $\epsilon_y/10^{-6}\text{mrad}$ &   0.25 &    0.25 &   0.08 
                                      &   0.1  &    0.06 &   0.1  \\\hline
      $\beta^*_x/\text{mm}$           &  10.98 &   24.95 &   8.00 
                                      &  32    &   25    &  10.00 \\\hline
      $\beta^*_y/\text{mm}$           &   0.45 &    0.70 &   0.13 
                                      &   0.8  &    0.7  &   0.12 \\\hline
      $\sigma_x/\text{nm}$            & 335    &  845    & 286    
                                      & 571.87 &  598.08 & 226    \\\hline
      $\sigma_y/\text{nm}$            &  15.1  &   18.9  &   4.52 
                                      &   9.04 &    6.55 &   3.57 \\\hline
      $\sigma_z/\mu\text{m}$          & 300    &  700    & 100    
                                      & 500    &  500    & 125    \\\hline
      $f_{\text{rep}}$                &  50    &    5    & 180    
                                      &  50    &    5    & 180    \\\hline
      $n_{\text{bunch}}$              & 333    & 1135    &  90    
                                      & 125    & 2270    &  90    \\\hline
    \end{tabular}
  \end{center}
  \caption{\label{tab:acc_param}%
    Accelerator parameters for three typical designs
    at~$\protect\sqrt s = 500\text{GeV}$
    and~$\protect\sqrt s = 1\text{TeV}$.   The resulting
    distributions are shown in figure~\ref{fig:dist}.  The
    design efforts are currently concentrated on a
    $350\text{GeV}$-$800\text{GeV}$ LC.  Therefore the Tesla
    parameters for~$1\text{TeV}$ are slightly out of date.}
\end{table}
\begin{figure}[tp]
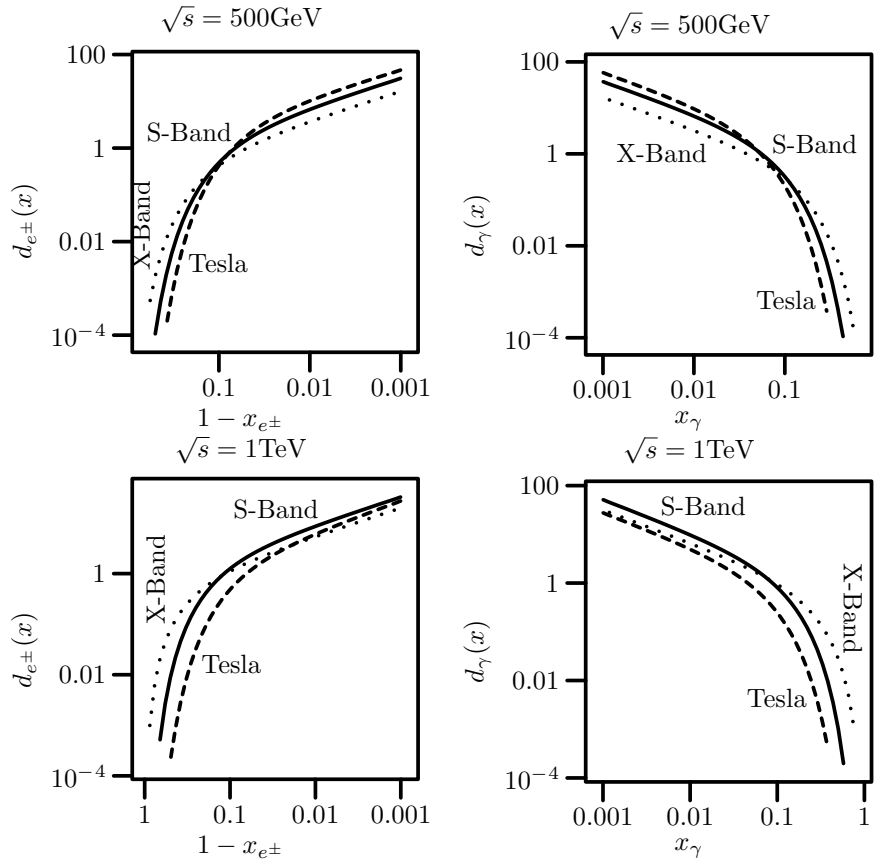

  \begin{center}
    \includegraphics{dist.1}\quad\includegraphics{dist.2} \\
    \includegraphics{dist.3}\quad\includegraphics{dist.4}
  \end{center}
  \caption{\label{fig:dist}%
    Version 1, revision 1996 07 11 of the factorized $e^\pm$- and
    $\gamma$-distributions at~$\protect\sqrt s = 500\text{GeV}$
    and~$\protect\sqrt s = 1\text{TeV}$ in a doubly logarithmic plot.
    The accelerator parameters are taken from
    table~\ref{tab:acc_param}.}
\end{figure}
%%%%%%%%%%%%%%%%%%%%%%%%%%%%%%%%%%%%%%%%%%%%%%%%%%%%%%%%%%%%%%%%%%%%%%%%
\begin{table}
  \begin{center}
    \renewcommand{\arraystretch}{1.3}
    \begin{tabular}{|c||c|c|c|}\hline
          & \texttt{TESLA} & \texttt{TESLA} & \texttt{TESLA}
      \\\hline\hline
      $E/\text{GeV}$                  &  175    &  250    &  400    \\\hline
      $N_{\text{particles}}/10^{10}$  &    3.63 &    3.63 &    3.63 \\\hline
      $\epsilon_x/10^{-6}\text{mrad}$ &   14    &   14    &   14    \\\hline
      $\epsilon_y/10^{-6}\text{mrad}$ &    0.25 &    0.25 &    0.1  \\\hline
      $\beta^*_x/\text{mm}$           &   25.00 &   24.95 &   15.00 \\\hline
      $\beta^*_y/\text{mm}$           &    0.70 &    0.70 &    0.70 \\\hline
      $\sigma_x/\text{nm}$            & 1010.94 &  845    &  668.67 \\\hline
      $\sigma_y/\text{nm}$            &   22.6  &   18.9  &    9.46 \\\hline
      $\sigma_z/\mu\text{m}$          &  700    &  700    &  700    \\\hline
      $f_{\text{rep}}$                &    5    &    5    &    5    \\\hline
      $n_{\text{bunch}}$              & 1135    & 1135    & 1135    \\\hline
    \end{tabular}
  \end{center}
  \caption{\label{tab:acc_param/Tesla}%
    Accelerator parameters for the Tesla design at three
    planned~\protect\cite{Tesla:1996:CDR} energies.  The resulting
    distributions are shown in figure~\ref{fig:dist/Tesla}.}
\end{table}
\begin{figure}[tp]
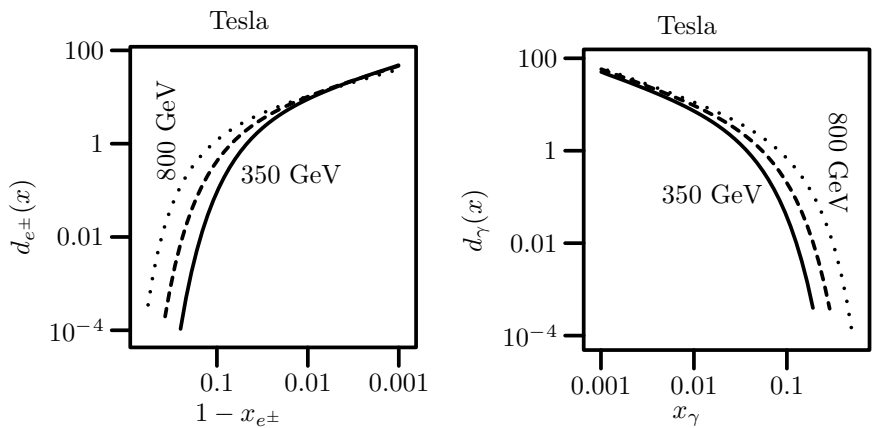

  \begin{center}
    \includegraphics{dist.5}\quad\includegraphics{dist.6}
  \end{center}
  \caption{\label{fig:dist/Tesla}%
    Version 1, revision 1996 07 11 of the factorized $e^\pm$- and
    $\gamma$-distributions for Tesla in a doubly logarithmic plot.
    The accelerator parameters are taken from
    table~\ref{tab:acc_param/Tesla}.}
\end{figure}
%%%%%%%%%%%%%%%%%%%%%%%%%%%%%%%%%%%%%%%%%%%%%%%%%%%%%%%%%%%%%%%%%%%%%%%%
The microscopic simulation program
\texttt{Guinea-Pig}~\cite{Schulte:1996:Thesis}
used for the current version of
the parameterizations in \Kirke/ simulates the
passage of electrons through a bunch of electrons (and vice versa).
It takes the following accelerator parameters as input:
\begin{description}
  \item[$E$]: the energy of the particles before the beam-beam interaction.
  \item[$N_{\text{particles}}$]: the number of particles per bunch.
  \item[$\epsilon_{x,y}$]: the normalized horizontal and vertical emittances.
  \item[$\beta^*_{x,y}$]: the horizontal and vertical beta functions.
  \item[$\sigma_{x,y,z}$]: the horizontal, vertical and longitudinal beam
    size.  A Gaussian shape is used for the charge distribution in the
    bunches. 
  \item[$f_{\text{rep}}$]: the repetition rate.
  \item[$n_{\text{bunch}}$]: the number of bunches per train.
\end{description}
The transversal beam sizes, beta functions and normalized emittances
for relativistic particles are related by
\begin{equation}
  \beta^*_{x,y} = \frac{\sigma_{x,y}^2}{\epsilon_{x,y}} \frac{E}{m_e}
\end{equation}
The parameters used in the most recent revision of the
parameterizations are collected in tables~\ref{tab:acc_param}
and~\ref{tab:acc_param/Tesla}.  The resulting
factorized electron/positron and photon distributions in version 1 of
the parameterizations are depicted in figures~\ref{fig:dist}
and~\ref{fig:dist/Tesla}.\par
The most important purpose of \Kirke/ is to map the manifold of
possible beam spectra for the NLC to a \emph{finite} number of
\emph{reproducible} parameterizations.  The distributions
\begin{equation}
\label{eq:dist}
  D^{\alpha\nu\rho}_{p_1p_2} (x_1, x_2; \sqrt s)
\end{equation}
provided by
\Kirke/ are indexed by three integers
\begin{description}
  \item[$\alpha$]: the \emph{accelerator design class:} currently
    there are three options: S-band~\cite{S-Band:1996:CDR},
    Tesla~\cite{Tesla:1996:CDR},
    X-band~\cite{JLC:1992:CDR,NLC:1996:ZDR}.  More
    variety will be added later, in particular the~$e^-e^-$ mode and
    the $e^-\gamma$ and~$\gamma\gamma$ laser backscattering modes of
    these designs.
  \item[$\nu$]: the \emph{version of the parameterization:} over the years,
    the form of the parameterizations can change, either because better
    approximations are found or because new simulation programs become
    available.  All versions will remain available in order to be able
    to reproduce calculations.
  \item[$\rho$]: the \emph{revision date for the parameterization:}  a
    particular parameterization can contain bugs, which will be fixed
    in subsequent revisions.  While only the most recent revision should
    be used for new calculations, old revisions will remain available
    in order to be able to reproduce calculations.
\end{description}
The continuous parameter~$\sqrt s$ in~(\ref{eq:dist}) is misleading, because
accelerator parameters have been optimized for discrete values of the
energy.  Therefore the distributions are not available for all values
of~$\sqrt s$.\par
The usage of the distributions in application programs is discussed in
section~\ref{sec:dist-usage}.
\Kirke/ provides for each of the distributions a non-uniform random
variate generator, that generates energy
fractions according to the distributions.  The usage of these
generators is discussed in section~\ref{sec:mc-usage}.
\nwenddocs{}%
\nwbegindocs{3}%
%%%%%%%%%%%%%%%%%%%%%%%%%%%%%%%%%%%%%%%%%%%%%%%%%%%%%%%%%%%%%%%%%%%%%%
\section{Usage}
\label{sec:usage}
\nwenddocs{}%
\nwbegindocs{4}%
%%%%%%%%%%%%%%%%%%%%%%%%%%%%%%%%%%%%%%%%%%%%%%%%%%%%%%%%%%%%%%%%%%%%%%
\subsection{Distributions}
\label{sec:dist-usage}
A generic interface to all distributions~$D_{p_1p_2}(x_1,x_2)$ is
given by the \texttt{circe} function
\nwenddocs{}%
\nwbegincode{5}\moddef{API documentation}%
\endmoddef
      double precision circe, d, x1, x2
      integer p1, p2
      d = circe (x1, x2, p1, p2)
\nwendcode{}%
\nwbegindocs{6}%
where the energy fractions are specified by~$x_{1,2}$ and the
particles~$p_{1,2}$ are identified by their standard Monte Carlo
codes:\cite{PDG:1994}
\nwenddocs{}%
\nwbegincode{7}\moddef{Particle codes}%
\endmoddef
      integer ELECTR, POSITR, PHOTON
      parameter (ELECTR =  11)
      parameter (POSITR = -11)
      parameter (PHOTON =  22)
\nwendcode{}%
\nwbegindocs{8}%
\nwdocspar
\nwenddocs{}%
\nwbegindocs{9}%
The distributions can have integrable singularities at the end points,
therefore the calling functions \emph{must not} evaluate them at the
endpoints~$0$ and~$1$. This is usually not a problem, since standard
mapping techniques~(cf.~(\ref{eq:mapping}) below) will have to be used
to take care of the singularity anyway.  Nevertheless, all
applications should favor open quadrature formulae (i.e.~formulae not
involving the endpoints) over closed formulae.
\nwenddocs{}%
\nwbegindocs{10}%
The distributions are guaranteed to vanish unless~$0<x_{1,2}<1$,
with two exceptions.  Firstly, the value~$-1$ allows to pick up the
integral of the continuum contribution:
\begin{subequations}
\begin{align}
  D_{p_1p_2}(-1,x_2) &=
    \lim_{\epsilon\to+0}\int_\epsilon^{1-\epsilon}\!dx_1\,
    D_{p_1p_2}(x_1,x_2) \\
  D_{p_1p_2}(x_1,-1) &=
    \lim_{\epsilon\to+0}\int_\epsilon^{1-\epsilon}\!dx_2\,
    D_{p_1p_2}(x_1,x_2) \\
  D_{p_1p_2}(-1,-1)  &=
    \lim_{\epsilon\to+0}\int_\epsilon^{1-\epsilon}\!dx_1dx_2\,
    D_{p_1p_2}(x_1,x_2)
\end{align}
\end{subequations}
The other exception is that the strength of $\delta$-function
contributions at the endpoint can be picked up from the value at this
endpoint:
\begin{subequations}
\begin{align}
  D_{e^+e^-} (x_1,x_2)
   &= D_{e^+e^-} (1,1) \delta(1-x_1) \delta(1-x_2)
       + \text{smooth and single $\delta$} \\
  D_{e^\pm\gamma} (x_1,x_2)
   &= D_{e^\pm\gamma} (1,x_2) \delta(1-x_1) + \text{smooth} \\
  D_{\gamma e^\pm} (x_1,x_2)
   &= D_{\gamma e^\pm} (x_1,1) \delta(1-x_2) + \text{smooth}
\end{align}
\end{subequations}
The use of these special values is demonstrated in an example in
section~\ref{sec:sample-int} below.\par
The distributions are normalized such that
\begin{equation}
  \lim_{\epsilon\to+0}
    \int_{-\epsilon}^{1+\epsilon}\!dx_1dx_2\, D_{e^+e^-}(x_1,x_2) = 1.
\end{equation}
and the nominal $e^+e^-$-luminosity of the currently active
accelerator design
can be retrieved from the database with the subroutine {\tt{}circel}.
The value is given in units of
\begin{equation}
  \text{fb}^{-1}\upsilon^{-1} = 10^{-32} \text{cm}^{-2}\text{sec}^{-1}
\end{equation}
where $\upsilon=10^7\text{sec}\approx \text{year}/\pi$ is an
``effective year'' of running with about 30\%\ up-time.
\nwenddocs{}%
\nwbegincode{11}\moddef{API documentation}%
\plusendmoddef
      double precision lumi
      call circel (lumi)
\nwendcode{}%
\nwbegindocs{12}%
A particular parameterization is selected by the {\tt{}circes} function:
\nwenddocs{}%
\nwbegincode{13}\moddef{API documentation}%
\plusendmoddef
      double precision x1m, x2m, roots
      integer acc, ver, rev, chat
      call circes (x1m, x2m, roots, acc, ver, rev, chat)
\nwendcode{}%
\nwbegindocs{14}%
The parameter {\tt{}roots} corresponds to the nominal center of mass
energy~$\sqrt s/\text{GeV}$ of the collider.  Currently
$\sqrt s = 350\text{GeV}, 500\text{GeV}, 800\text{GeV}, 1\text{TeV}$
(i.e.~{\tt{}350D0}, {\tt{}500D0}, {\tt{}800D0} and~{\tt{}1000D0}) are supported.
Application programs can \emph{not} assume
that energy values are interpolated.  For convenience, e.g.~in top
threshold scans around~$350\text{GeV}$, a small interval around the
supported values will be accepted as synonymous with the central
value, but a warning will be printed.
Section~\ref{sec:parameterizations} should be consulted for the
discrete values supported by a particular version of the
parameterizations.  Negative values of {\tt{}roots} will keep the
currently active value for~$\sqrt s$.\par
The parameters {\tt{}x1m} and {\tt{}x2m} will set
thresholds~$x_{1,\text{min}}$ and~$x_{2,\text{min}}$ for the event
generation in the routines described in
section~\ref{sec:mc-usage}.\par
The parameter {\tt{}acc} selects the accelerator design.
Currently the following accelerator codes are recognized:
\nwenddocs{}%
\nwbegincode{15}\moddef{Accelerator codes}%
\endmoddef
      integer SBAND, TESLA, XBAND
      parameter (SBAND  =  1, TESLA  =  2, XBAND  =  3)
      integer NACC
      parameter (NACC = 3)
\nwendcode{}%
\nwbegindocs{16}%
\nwdocspar
\nwenddocs{}%
\nwbegindocs{17}%
Negative values will keep the currently active accelerator.
Later I will add the~$e^-e^-$ mode and the $e^-\gamma$
and~$\gamma\gamma$ laser backscattering modes of these designs:
\nwenddocs{}%
\nwbegincode{18}\moddef{Future API documentation}%
\endmoddef
      integer SBAND, TESLA, XBAND
      parameter (SBAND  =  1, TESLA  =  2, XBAND  =  3)
      integer SBNDEE, TESLEE, XBNDEE
      parameter (SBNDEE =  4, TESLEE =  5, XBNDEE =  6)
      integer SBNDEG, TESLEG, XBNDEG
      parameter (SBNDEG =  7, TESLEG =  8, XBNDEG =  9)
      integer SBNDGG, TESLGG, XBNDGG
      parameter (SBNDGG = 10, TESLGG = 11, XBNDGG = 12)
      integer NACC
      parameter (NACC = 12)
\nwendcode{}%
\nwbegindocs{19}%
The {\tt{}ver} parameter is used to determine the version as follows:
\begin{description}
  \item[$\text{{\tt{}ver}}>0$]: a frozen version which is documented in
    section~\ref{sec:parameterizations}. For example, version 1 is a
    family of factorized Beta distributions:
    $D(x_1,x_2) \propto x_1^{a_1}(1-x_1)^{b_1} x_2^{a_2}(1-x_2)^{b_2}$.
  \item[$\text{{\tt{}ver}}=0$]: the latest experimental version, which is usually
    not documented and can change at any time without announcement.
  \item[$\text{{\tt{}ver}}<0$]: keep the currently active version.
\end{description}
\nwenddocs{}%
\nwbegindocs{20}%
The {\tt{}rev} parameter is used to determine the revision of a version
as follows: 
\begin{description}
  \item[$\text{{\tt{}rev}}>0$]: a frozen revision which is documented in
    section~\ref{sec:parameterizations}.  The integer {\tt{}rev} is
    constructed from the date as follows: $\text{{\tt{}rev}} =
    10^4\cdot\text{year} + 10^2\cdot\text{month} + \text{day}$, where
    the year is greater than 1995. Since Fortran77 ignores whitespace,
    it can be written like \verb+1996 07 11+ for readability.  If
    there is no exact match, the most recent revision before the
    specified date is chosen.
  \item[$\text{{\tt{}rev}}=0$]: the most recent revision.
  \item[$\text{{\tt{}rev}}<0$]: keep the currently active revision.
\end{description}
\nwenddocs{}%
\nwbegindocs{21}%
Finally, the parameter {\tt{}chat} controls the ``chattiness'' of
{\tt{}circe}.  If it is~$0$, only error messages are printed. If it is~$1$,
the parameters in use are printed whenever they change.  Higher
values of {\tt{}chat} can produce even more diagnostics.\par
In addition to the generic interface {\tt{}circe}, there are specialized
functions for particular particle distributions.  Obviously
\begin{equation}
   D_{e^\pm\gamma}^{\alpha\nu\rho} (x_1, x_2, s)
     = D_{\gamma e^\pm}^{\alpha\nu\rho} (x_2, x_1, s)
\end{equation}
and there are three independent functions~$D_{e^-e^+}$,
$D_{e^-\gamma}$ and~$D_{\gamma\gamma}$ for the~$e^+e^-$ colliders with
reasonable mnemonics:
\nwenddocs{}%
\nwbegincode{22}\moddef{API documentation}%
\plusendmoddef
      double precision circee, circeg, circgg
      d = circee (x1, x2)
      d = circeg (x1, x2)
      d = circgg (x1, x2)
\nwendcode{}%
\nwbegindocs{23}%
Calling the latter three functions is marginally faster in the
current implementation, but this can change in the future.
\nwenddocs{}%
\nwbegindocs{24}%
%%%%%%%%%%%%%%%%%%%%%%%%%%%%%%%%%%%%%%%%%%%%%%%%%%%%%%%%%%%%%%%%%%%%%%
\subsubsection{Example}
\label{sec:sample-int}
For clarification, let me give a simple example.  Imagine we want to
calculate the integrated production cross section
\begin{equation}
  \sigma_X(s) = \int\!dx_1dx_2\,
     \sigma_{e^+e^-\to X}(x_1x_2s) D_{e^+e^-}(x_1,x_2,s)
\end{equation}
Since the distributions are singular in the~$x_{1,2}\to1$ limit, we
have to map away this singularity with
\begin{subequations}
\label{eq:mapping}
\begin{align}
  x &\to t = (1-x)^{1/\eta}\\
\intertext{Therefore}
  \int_0^1\!dx\,f(x) &= \int_0^1\!dt\, \eta t^{\eta-1} f(1-t^\eta)
\end{align}
\end{subequations}
with~$\eta$ sufficiently large to give the integrand a finite limit
at~$x\to1$.  If~$f$ diverges like a power~$f(x) \propto 1/(1-x)^\beta$,
this means~$\eta>1/(1-\beta)$.\par
As a specific example, let us ``measure'' a one particle $s$-channel
exchange cross section
\begin{equation}
  \sigma(s) \propto \frac{1}{s}
\end{equation}
\nwenddocs{}%
\nwbegincode{25}\moddef{\code{}sample.f\edoc{}}%
\endmoddef
      double precision function sigma (s)
      implicit none
      double precision s
      sigma = 1d0 / s
      end
\nwendcode{}%
\nwbegindocs{26}%
I will present the example code in a bottom-up fashion, which
should be intuitive and is described in some more detail in
appendix~\ref{sec:litprog}. 
Assuming the existence of a one- and a two-dimensional Gaussian
integration function {\tt{}gauss1} and {\tt{}gauss2},\footnote{%
  They are provided in the example program {\tt{}sample.f}.}
we can perform the integral as follows:
\nwenddocs{}%
\nwbegincode{27}\moddef{Gauss integration}%
\endmoddef
      s = sigma (1d0) * circee (1d0, 1d0)
     $   + gauss1 (d1, 0d0, 1d0, EPS)
     $   + gauss1 (d2, 0d0, 1d0, EPS)
     $   + gauss2 (d12, 0d0, 1d0, 0d0, 1d0, EPS)
      write (*, 1000) 'delta(sigma) (Gauss) =', (s-1d0)*100d0
 1000 format (1X, A22, 1X, F5.2, '%')
\nwendcode{}%
\nwbegindocs{28}%
Note how the four combinations of continuum and $\delta$-peak are
integrated separately, where you have to use three auxiliary
functions~{\tt{}d1}, {\tt{}d2} and~{\tt{}d12}.
The continuum contribution, including the Jacobian:
\nwenddocs{}%
\nwbegincode{29}\moddef{\code{}sample.f\edoc{}}%
\plusendmoddef
      double precision function d12 (t1, t2)
      implicit none
      double precision t1, t2, x1, x2, sigma, circee
      \LA{}\code{}EPS\edoc{} \&\ \code{}PWR\edoc{}~{\nwtagstyle{}}\RA{}
      x1 = 1d0 - t1**PWR
      x2 = 1d0 - t2**PWR
      d12 = PWR*PWR * (t1*t2)**(PWR-1d0)
     $       * sigma (x1*x2) * circee (x1, x2)
      end
\nwendcode{}%
\nwbegindocs{30}%
the first product of continuum and $\delta$-peak:
\nwenddocs{}%
\nwbegincode{31}\moddef{\code{}sample.f\edoc{}}%
\plusendmoddef
      double precision function d1 (t1)
      implicit none
      double precision t1, x1, sigma, circee
      \LA{}\code{}EPS\edoc{} \&\ \code{}PWR\edoc{}~{\nwtagstyle{}}\RA{}
      x1 = 1d0 - t1**PWR
      d1 = PWR * t1**(PWR-1d0) * sigma (x1) * circee (x1, 1d0)
      end
\nwendcode{}%
\nwbegindocs{32}%
and the second one:
\nwenddocs{}%
\nwbegincode{33}\moddef{\code{}sample.f\edoc{}}%
\plusendmoddef
      double precision function d2 (t2)
      implicit none
      double precision t2, x2, sigma, circee
      \LA{}\code{}EPS\edoc{} \&\ \code{}PWR\edoc{}~{\nwtagstyle{}}\RA{}
      x2 = 1d0 - t2**PWR
      d2 = PWR * t2**(PWR-1d0) * sigma (x2) * circee (1d0, x2)
      end
\nwendcode{}%
\nwbegindocs{34}%
Below you will see that the power of the singularity of the
$e^+e^-$~distributions at~$x\to1$ is~$\approx-2/3$.  To be on the safe
side, we choose the power~$\eta$ in~(\ref{eq:mapping}) as~$5$.  It is
kept in the parameter~{\tt{}PWR}, while~{\tt{}EPS} is the desired accuracy
of the Gaussian integration:
\nwenddocs{}%
\nwbegincode{35}\moddef{\code{}EPS\edoc{} \&\ \code{}PWR\edoc{}}%
\endmoddef
      double precision EPS, PWR
      parameter (EPS = 1d-6, PWR = 5d0)
\nwendcode{}%
\nwbegindocs{36}%
These code fragments can now be used in a main program that loops
over energies and accelerator designs
\nwenddocs{}%
\nwbegincode{37}\moddef{\code{}sample.f\edoc{}}%
\plusendmoddef
      program sample
      implicit none
      \LA{}Accelerator codes~{\nwtagstyle{}}\RA{}
      \LA{}\code{}EPS\edoc{} \&\ \code{}PWR\edoc{}~{\nwtagstyle{}}\RA{}
      \LA{}Other variables in \code{}sample\edoc{}~{\nwtagstyle{}}\RA{}
      integer acc, ver, i
      double precision roots(2)
      data roots / 500D0, 1000D0 /
      do 10 acc = 1, NACC
         do 11 ver = 1, 1
            do 12 i = 1, 2
               call circes (0d0, 0d0, roots(i), acc, ver, 1996 07 29, 1)
               \LA{}Gauss integration~{\nwtagstyle{}}\RA{}
               \LA{}Monte Carlo integration~{\nwtagstyle{}}\RA{}
 12         continue
 11      continue
 10   continue
      end
\nwendcode{}%
\nwbegindocs{38}%
with the following result
\nwenddocs{}%
\nwbegincode{39}\moddef{Sample output}%
\endmoddef
 circe:message: starting up ...
 circe:message: Id: circe.nw,v 1.22 1996/07/27 19:52:28 ohl Exp
 circe:message: updating `acc' to SBAND                                     
 circe:message: updating `ver' to  1                                        
 circe:message: updating `rev' to 19960729                                  
 delta(sigma) (Gauss) =  3.79%
 delta(sigma) (MC)    =  3.74%
                    +/-   .06%
 circe:message: updating `roots' to 1000.0                                  
 delta(sigma) (Gauss) = 10.11%
 delta(sigma) (MC)    =  9.97%
                    +/-   .14%
 circe:message: updating `roots' to  500.0                                  
 circe:message: updating `acc' to TESLA                                     
 delta(sigma) (Gauss) =  3.11%
 delta(sigma) (MC)    =  3.09%
                    +/-   .04%
 circe:message: updating `roots' to 1000.0                                  
 delta(sigma) (Gauss) =  3.98%
 delta(sigma) (MC)    =  3.96%
                    +/-   .07%
 circe:message: updating `roots' to  500.0                                  
 circe:message: updating `acc' to XBAND                                     
 delta(sigma) (Gauss) =  4.96%
 delta(sigma) (MC)    =  4.96%
                    +/-   .10%
 circe:message: updating `roots' to 1000.0                                  
 delta(sigma) (Gauss) = 21.31%
 delta(sigma) (MC)    = 21.72%
                    +/-   .45%
\nwendcode{}%
\nwbegindocs{40}%
We almost forgot to declare the variables in the main program
\nwenddocs{}%
\nwbegincode{41}\moddef{Other variables in \code{}sample\edoc{}}%
\endmoddef
      double precision s
      double precision gauss1, gauss2, circee, sigma, d1, d2, d12
      external d1, d2, d12
\nwendcode{}%
\nwbegindocs{42}%
This concludes the integration example.  It should have made it
obvious how to proceed in a realistic application.\par
In section~\ref{sec:sample-MC} below, I will describe a Monte Carlo
method for calculating such integrals efficiently.
\nwenddocs{}%
\nwbegindocs{43}%
%%%%%%%%%%%%%%%%%%%%%%%%%%%%%%%%%%%%%%%%%%%%%%%%%%%%%%%%%%%%%%%%%%%%%%
\subsection{Generators}
\label{sec:mc-usage}
The function {\tt{}circe} and its companions are opaque to the
user. Since they will in general contain singularities, applications
will \emph{not} be able to generate corresponding samples of random
numbers efficiently.  To fill this gap, four random number generators
are provided.  The subroutine {\tt{}girce} will generate particle
types~$p_{1,2}$ and energy fractions~$x_{1,2}$ in one step, according
to the selected distribution.\footnote{\index{inefficiencies}% 
  The implementation of the flavor selection with non-vanishing
  thresholds~$x_{1,\text{min}}$ and~$x_{2,\text{min}}$ is moderately
  inefficient at the moment.  It can be improved by a factor of two.}
Particle~$p_1$ will be either a positron or a photon and~$p_2$ will be
either an electron or a photon.  The energy fractions are guaranteed
to be above the currently active thresholds: $x_i \ge
x_{i,\text{min}}$.  This can be used to cut on soft events---the
photon distributions are rather soft---which might not be interesting
in most simulations.
\nwenddocs{}%
\nwbegincode{44}\moddef{API documentation}%
\plusendmoddef
      call girce  (x1, x2, p1, p2, rng)
\nwendcode{}%
\nwbegindocs{45}%
The output parameters of {\tt{}girce} are identical to the input
parameters of {\tt{}circe}, with the exception of {\tt{}rng}.  The latter is
a subroutine with a single double precision argument, 
which will be assigned a uniform deviate from the interval $[0,1]$
after each call:
\nwenddocs{}%
\nwbegincode{46}\moddef{API documentation}%
\plusendmoddef
      subroutine rng (r)
      double precision r
      r = \LA{}uniform deviate on $[0,1]$~{\nwtagstyle{}}\RA{}
      end
\nwendcode{}%
\nwbegindocs{47}%
Typically, it will be just a wrapper around the standard random
number generator of the application program.
\nwenddocs{}%
\nwbegindocs{48}%
For studies with a definite initial state, three generator functions
are available.
\nwenddocs{}%
\nwbegincode{49}\moddef{API documentation}%
\plusendmoddef
      call gircee (x1, x2, rng)
      call girceg (x1, x2, rng)
      call gircgg (x1, x2, rng)
\nwendcode{}%
\nwbegindocs{50}%
%%%%%%%%%%%%%%%%%%%%%%%%%%%%%%%%%%%%%%%%%%%%%%%%%%%%%%%%%%%%%%%%%%%%%%
\subsubsection{Example}
\label{sec:sample-MC}
Returning to the example from section~\ref{sec:sample-MC}, I present
a concise Monte Carlo algorithm for calculating the same integral:
\nwenddocs{}%
\nwbegincode{51}\moddef{Monte Carlo integration}%
\endmoddef
      s = 0d0
      s2 = 0d0
      do 100 n = 1, NEVENT
         call gircee (x1, x2, random)
         w = sigma (x1*x2)
         s = s + w
         s2 = s2 + w*w
 100  continue
      s = s / dble(NEVENT)
      s2 = s2 / dble(NEVENT)
      write (*, 1000) 'delta(sigma) (MC)    =', (s-1d0)*100d0
      write (*, 1000) '                   +/-',
     $                sqrt((s2-s*s)/dble(NEVENT))*100d0
\nwendcode{}%
\nwbegindocs{52}%
\nwdocspar
\nwenddocs{}%
\nwbegincode{53}\moddef{Other variables in \code{}sample\edoc{}}%
\plusendmoddef
      double precision w, s2, x1, x2
      external random
      integer NEVENT, n
      parameter (NEVENT = 10000)
\nwendcode{}%
\nwbegindocs{54}%
Here is a simple linear congruential random number generator for the
sample program.  Real applications will use their more sophisticated
generators instead.
\nwenddocs{}%
\nwbegincode{55}\moddef{\code{}sample.f\edoc{}}%
\plusendmoddef
      subroutine random (r)
      implicit none
      double precision r
      integer m, a, c
      parameter (M = 259200, A = 7141, C = 54773)
      integer n
      save n
      data n /0/
      n = mod(n*a+c,m)
      r = dble (n) / dble (m)
      end
\nwendcode{}%
\nwbegindocs{56}%
If the cross section is slowly varying on the range where
the~$x_{1,2}$ distributions are non-zero, this algorithm is very
efficient.\par
However, if this condition is not met, the explicit form of the
parameterizations in section~\ref{sec:parameterizations} should be
consulted and appropriate mapping techniques should be applied.  The
typical example for this problem is a narrow resonance just below the
nominal beam energy.
\nwenddocs{}%
\nwbegindocs{57}%
%%%%%%%%%%%%%%%%%%%%%%%%%%%%%%%%%%%%%%%%%%%%%%%%%%%%%%%%%%%%%%%%%%%%%%
\subsubsection{Event Generators}
\label{sec:MC}
For Monte Carlo event generators that use the standard {\tt{}/hepevt/}
common block~\cite{Altarelli/etal:1989:LEP1}, the addition of the
\Kirke/ library is trivial.  During the initialization of the event
generator, the {\tt{}circes} subroutine is called to set up \Kirke/'s
internal state.  For example:
\nwenddocs{}%
\nwbegincode{58}\moddef{Initialize event generator}%
\endmoddef
      call circes (0d0, 0d0, roots, acc, ver, 1996 07 11, 1)
\nwendcode{}%
\nwbegindocs{59}%
During event generation, before setting up the~$e^+e^-$ initial
state, the {\tt{}gircee} subroutine is called with the event generator's
random number generator:
\nwenddocs{}%
\nwbegincode{60}\moddef{Event generation}%
\endmoddef
      call gircee (x1, x2, random)
\nwendcode{}%
\nwbegindocs{61}%
The resulting energy fractions~$x_1$ and~$x_2$ are now available for
defining the initial state electron
\nwenddocs{}%
\nwbegincode{62}\moddef{Event generation}%
\plusendmoddef
      isthep(1) = 101
      idhep(1) = ELECTR
      phep(1,1) = 0d0
      phep(2,1) = 0d0
      phep(3,1) = x1 * ebeam
      phep(4,1) = x1 * ebeam
      phep(5,1) = 0d0
\nwendcode{}%
\nwbegindocs{63}%
and positron.
\nwenddocs{}%
\nwbegincode{64}\moddef{Event generation}%
\plusendmoddef
      isthep(2) = 102
      idhep(2) = POSITR
      phep(1,2) = 0d0
      phep(2,2) = 0d0
      phep(3,2) = - x2 * ebeam
      phep(4,2) = x2 * ebeam
      phep(5,2) = 0d0
\nwendcode{}%
\nwbegindocs{65}%
Using \Kirke/ with other event generators should be straightforward
as well.
\nwenddocs{}%
\nwbegindocs{66}%
%%%%%%%%%%%%%%%%%%%%%%%%%%%%%%%%%%%%%%%%%%%%%%%%%%%%%%%%%%%%%%%%%%%%%%
\section{Technical Notes}
\label{sec:technical}
\begin{figure}[tp]
  \begin{center}
    \includegraphics{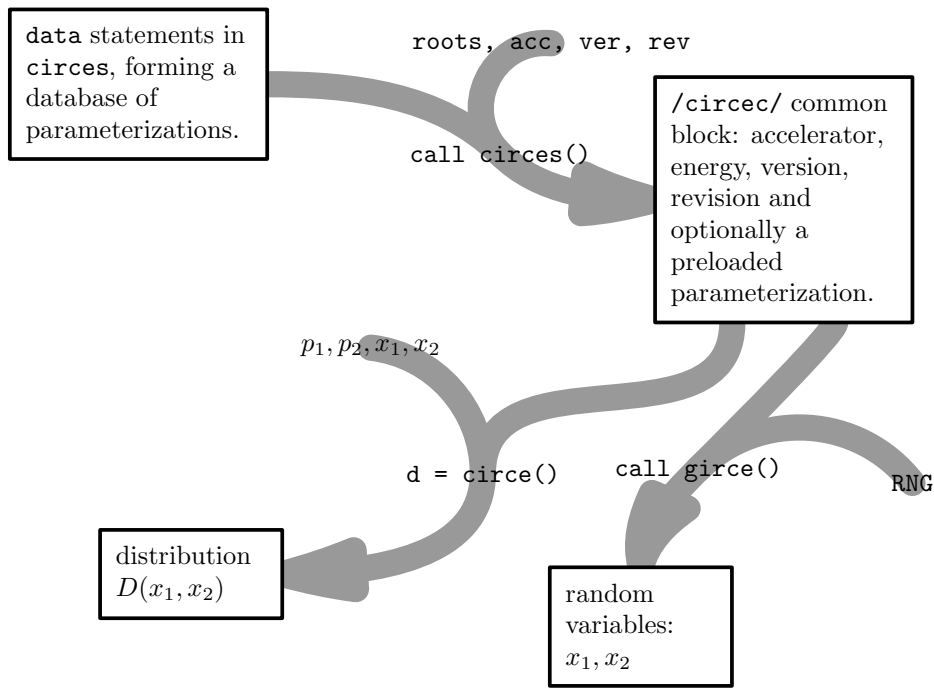}
  \end{center}
  \caption{\label{fig:architecture}%
    Architecture of \Kirke/: {\tt{}circes()} selects energy and accelerator
    and loads the parameterization.  The function {\tt{}circe()}
    calculates the values of the selected distribution function at the
    given energy fractions.  The subroutine {\tt{}girce()} generates
    energy fractions using a specified random number generator in
    accordance with the selected distribution.}
\end{figure}
The structure of \Kirke/ is extremely
simple~(cf.~figure~\ref{fig:architecture}) and is mainly a bookkeeping
excercise.  All that needs to be done is to maintain a database of
available parameterizations and to evaluate the corresponding
functions. The only non trivial algorithms are used for the efficient
generation of random deviates.\par
I have avoided the use of initialized \texttt{common} blocks
(i.e.~\texttt{block data} subroutines), because the Fortran77 standard
does not provide a \emph{portable} way of ensuring that \texttt{block
data} subroutines are actually executed at loading time.  Instead,
the \texttt{/circom/} common block is tagged by a ``magic number'' to
check for initialization and its members are filled by the
\texttt{circes} subroutine when necessary.\par
A more flexible method would be to replace the \texttt{data}
statements by reading external files.  This option causes portability
problems, however, because I would have to make sure that the names
of the external files are valid in all files systems of the target
operating systems.  More significantly, splitting the implementation
into several parts forces the user to keep all files up to date.
This can be a problem, because Fortran source files and data input
files will typically be kept in different parts of the file
system.\par
The option of implementing \Kirke/ statelessly, i.e.~with pure
function calls and without \texttt{common} blocks, has been
dismissed.  While it would have been more straightforward on the side
of the library, it would have placed the burdon of maintaining state
(accelerator, energy, etc.) on the application program, thereby
complicating them considerably.  Keeping an explicit state in \Kirke/
has the additional benefit of allowing to precompute certain internal
variables, resulting in a more efficient implementation.
\nwenddocs{}%
\nwbegindocs{67}%
%%%%%%%%%%%%%%%%%%%%%%%%%%%%%%%%%%%%%%%%%%%%%%%%%%%%%%%%%%%%%%%%%%%%%%
\section{Parameterizations}
\label{sec:parameterizations}
Version \Version/ of \Kirke/ supports just one version of the
parameterizations.  Future versions will provide additional
parameterizations. 
\nwenddocs{}%
\nwbegindocs{68}%
%%%%%%%%%%%%%%%%%%%%%%%%%%%%%%%%%%%%%%%%%%%%%%%%%%%%%%%%%%%%%%%%%%%%%%
\subsection{Version 1}
\label{sec:factorized-beta}
%%%%%%%%%%%%%%%%%%%%%%%%%%%%%%%%%%%%%%%%%%%%%%%%%%%%%%%%%%%%%%%%%%%%%%%%
\begin{table}
  \begin{center}
    \renewcommand{\arraystretch}{1.3}
    \begin{tabular}{|c||c|c|c|}\hline
      & \texttt{SBAND} & \texttt{TESLA} & \texttt{XBAND}
      \\\hline\hline
$\mathcal{L}/\text{fb}^{-1}\upsilon^{-1}$
 & $  31.47_{-0.12}^{+0.12}$
 & $ 106.08_{-0.38}^{+0.38}$
 & $  36.15_{-0.16}^{+0.16}$
\\\hline
$\int d_{e^\pm}$
 & $  .6170_{-0.0029}^{+0.0029}$
 & $  .7172_{-0.0033}^{+0.0033}$
 & $  .4872_{-0.0028}^{+0.0029}$
\\\hline
$x_{e^\pm}^\alpha$
 & $12.6180_{-0.0423}^{+0.0426}$
 & $19.2577_{-0.0539}^{+0.0541}$
 & $ 7.5135_{-0.0360}^{+0.0364}$
\\\hline
$(1-x_{e^\pm})^\alpha$
 & $ -.6161_{-0.0007}^{+0.0007}$
 & $ -.5839_{-0.0007}^{+0.0007}$
 & $ -.6225_{-0.0009}^{+0.0009}$
\\\hline
$\int d_\gamma$
 & $  .6378_{-0.0019}^{+0.0019}$
 & $  .7593_{-0.0021}^{+0.0021}$
 & $  .4306_{-0.0019}^{+0.0019}$
\\\hline
$x_\gamma^\alpha$
 & $ -.6896_{-0.0004}^{+0.0004}$
 & $ -.6940_{-0.0003}^{+0.0003}$
 & $ -.6853_{-0.0006}^{+0.0006}$
\\\hline
$(1-x_\gamma)^\alpha$
 & $15.0658_{-0.0444}^{+0.0446}$
 & $23.6384_{-0.0628}^{+0.0630}$
 & $ 8.5519_{-0.0373}^{+0.0376}$
\\\hline
    \end{tabular}
  \end{center}
  \caption{\label{tab:param}%
    Version 1, revision 1996 07 11 of the beam spectra at 500 GeV.
    The rows correspond to the luminosity per effective year, the
    integral over the continuum and the powers in the factorized Beta
    distributions~(\ref{eq:beta}).}
\end{table}
\begin{table}
  \begin{center}
    \renewcommand{\arraystretch}{1.3}
    \begin{tabular}{|c||c|c|c|}\hline
      & \texttt{SBAND} & \texttt{TESLA} & \texttt{XBAND}
      \\\hline\hline
$\mathcal{L}/\text{fb}^{-1}\upsilon^{-1}$
 & $ 245.66_{-0.78}^{+0.78}$
 & $ 109.36_{-0.42}^{+0.42}$
 & $ 117.99_{-0.48}^{+0.48}$
\\\hline
$\int d_{e^\pm}$
 & $  .7599_{-0.0028}^{+0.0028}$
 & $  .5896_{-0.0030}^{+0.0030}$
 & $  .6876_{-0.0030}^{+0.0030}$
\\\hline
$x_{e^\pm}^\alpha$
 & $ 6.9085_{-0.0145}^{+0.0145}$
 & $11.6104_{-0.0439}^{+0.0441}$
 & $ 2.9938_{-0.0094}^{+0.0095}$
\\\hline
$(1-x_{e^\pm})^\alpha$
 & $ -.5515_{-0.0007}^{+0.0007}$
 & $ -.6124_{-0.0008}^{+0.0008}$
 & $ -.5585_{-0.0009}^{+0.0010}$
\\\hline
$\int d_\gamma$
 & $  .8216_{-0.0018}^{+0.0018}$
 & $  .4999_{-0.0019}^{+0.0019}$
 & $  .7275_{-0.0019}^{+0.0019}$
\\\hline
$x_\gamma^\alpha$
 & $ -.6862_{-0.0003}^{+0.0003}$
 & $ -.6907_{-0.0005}^{+0.0005}$
 & $ -.6712_{-0.0004}^{+0.0004}$
\\\hline
$(1-x_\gamma)^\alpha$
 & $ 9.4494_{-0.0208}^{+0.0208}$
 & $14.6981_{-0.0531}^{+0.0535}$
 & $ 4.1119_{-0.0111}^{+0.0111}$
\\\hline
    \end{tabular}
  \end{center}
  \caption{\label{tab:param/TeV}%
    Version 1, revision 1996 07 29 of the beam spectra at 1 TeV.}
\end{table}
\begin{table}
  \begin{center}
    \renewcommand{\arraystretch}{1.3}
    \begin{tabular}{|c||c|c|c|}\hline
      & 350 GeV & 500 GeV & 800 GeV
      \\\hline\hline
$\mathcal{L}/\text{fb}^{-1}\upsilon^{-1}$
 & $  74.70_{-0.28}^{+0.28}$
 & $ 106.08_{-0.38}^{+0.38}$
 & $ 289.11_{-0.94}^{+0.94}$
\\\hline
$\int d_{e^\pm}$
 & $  .6531_{-0.0033}^{+0.0033}$
 & $  .7172_{-0.0033}^{+0.0033}$
 & $  .7898_{-0.0031}^{+0.0031}$
\\\hline
$x_{e^\pm}^\alpha$
 & $33.7197_{-0.1084}^{+0.1089}$
 & $19.2577_{-0.0539}^{+0.0541}$
 & $ 9.6763_{-0.0194}^{+0.0195}$
\\\hline
$(1-x_{e^\pm})^\alpha$
 & $ -.5952_{-0.0007}^{+0.0007}$
 & $ -.5839_{-0.0007}^{+0.0007}$
 & $ -.5402_{-0.0007}^{+0.0007}$
\\\hline
$\int d_\gamma$
 & $  .6378_{-0.0022}^{+0.0022}$
 & $  .7593_{-0.0021}^{+0.0021}$
 & $  .8736_{-0.0019}^{+0.0019}$
\\\hline
$x_\gamma^\alpha$
 & $ -.6952_{-0.0004}^{+0.0004}$
 & $ -.6940_{-0.0003}^{+0.0003}$
 & $ -.6908_{-0.0003}^{+0.0003}$
\\\hline
$(1-x_\gamma)^\alpha$
 & $38.4884_{-0.1199}^{+0.1204}$
 & $23.6384_{-0.0628}^{+0.0630}$
 & $12.7329_{-0.0283}^{+0.0284}$
\\\hline
    \end{tabular}
  \end{center}
  \caption{\label{tab:param/Tesla}%
    Version 1, revision 1996 07 29 of the beam spectra for TESLA.}
\end{table}
%%%%%%%%%%%%%%%%%%%%%%%%%%%%%%%%%%%%%%%%%%%%%%%%%%%%%%%%%%%%%%%%%%%%%%%%
\begin{figure}[tp]
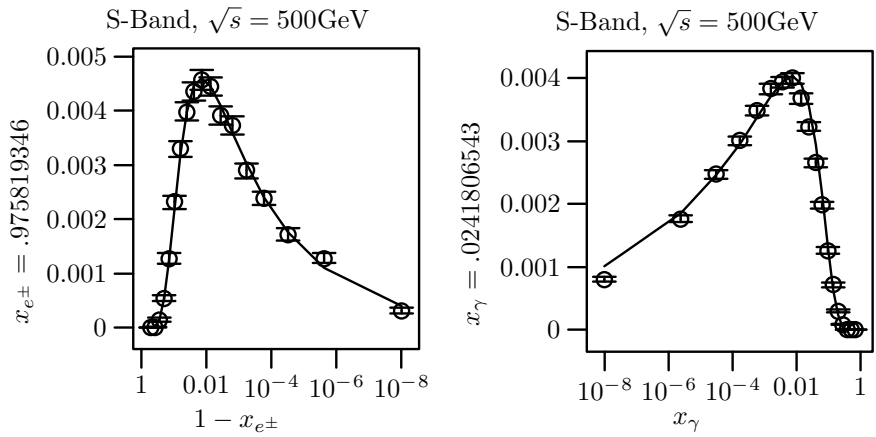

  \begin{center}
    \includegraphics{fit.11}\quad\includegraphics{fit.21}
  \end{center}
  \caption{\label{fig:fit/S-Band/500GeV}%
    Fit of the $e^\pm$- and $\gamma$-distributions for the S-Band
    design at $\protect\sqrt s = 500\text{GeV}$.  The open circles
    with error bars are the result of the \texttt{Guinea-Pig}
    similation. The full line is the fit.}
\end{figure}
\begin{figure}[tp]
  \begin{center}
    \includegraphics{fit.12}\quad\includegraphics{fit.22}
  \end{center}
  \caption{\label{fig:fit/Tesla/500GeV}%
    Fit of the $e^\pm$- and $\gamma$-distributions for the Tesla
    design at $\protect\sqrt s = 500\text{GeV}$.}
\end{figure}
\begin{figure}[tp]
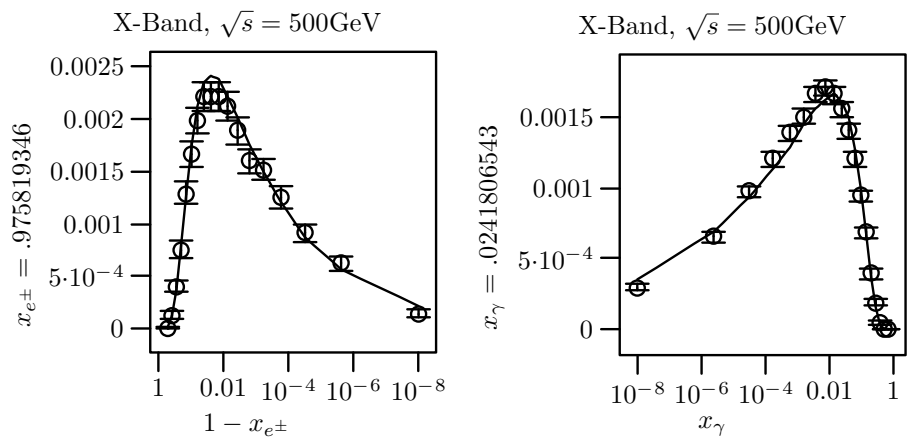

  \begin{center}
    \includegraphics{fit.13}\quad\includegraphics{fit.23}
  \end{center}
  \caption{\label{fig:fit/X-Band/500GeV}%
    Fit of the $e^\pm$- and $\gamma$-distributions for the X-Band
    design at $\protect\sqrt s = 500\text{GeV}$.}
\end{figure}
\begin{figure}[tp]
  \begin{center}
    \includegraphics{fit.15}\quad\includegraphics{fit.25}
  \end{center}
  \caption{\label{fig:fit/Tesla/1TeV}%
    Fit of the $e^\pm$- and $\gamma$-distributions for the Tesla
    design at $\protect\sqrt s = 1\text{TeV}$.}
\end{figure}
%%%%%%%%%%%%%%%%%%%%%%%%%%%%%%%%%%%%%%%%%%%%%%%%%%%%%%%%%%%%%%%%%%%%%%%%
The first version of the parameterization uses a simple factorized
\textit{ansatz}
\begin{subequations}
\label{eq:beta}
\begin{align}
  D_{p_1p_2}^{\alpha1\rho} (x_1,x_2,s)
   &= d_{p_1}^{\alpha1\rho} (x_1)  d_{p_2}^{\alpha1\rho} (x_2)\\
\intertext{where the distributions are simple Beta distributions:}
  d_{e^\pm}^{\alpha1\rho} (x)
   &= a_0^{\alpha\rho} \delta(1-x)
     + a_1^{\alpha\rho} x^{a_2^{\alpha\rho}} (1-x)^{a_3^{\alpha\rho}} \\
  d_\gamma^{\alpha1\rho} (x)
   &= a_4^{\alpha\rho} x^{a_5^{\alpha\rho}} (1-x)^{a_6^{\alpha\rho}}
\end{align}
\end{subequations}
This form of the distributions is motivated by the
observation~\cite{Chen/Noble:1986:Beamstrahlung}
that the $e^\pm$~distributions diverge like a power for $x\to1$ and
vanish at $x\to0$.  The behavior of the $\gamma$~distributions is
similar with the borders exchanged.
\nwenddocs{}%
\nwbegindocs{69}%
%%%%%%%%%%%%%%%%%%%%%%%%%%%%%%%%%%%%%%%%%%%%%%%%%%%%%%%%%%%%%%%%%%%%%%
\subsubsection{Fitting}
The parameters~$a_i$ in~(\ref{eq:beta}) have been obtained by a
least-square fit of~(\ref{eq:beta}) to histograms of simulation
results from \texttt{Guinea-Pig}.
Some care has to taken when fitting singular distributions to
histogrammed data.  Obviously equidistant bins are not a good idea,
because most bins will be almost empty (cf.~figures~\ref{fig:dist}
and~\ref{fig:dist/Tesla}) and consequently a lot of information will
be wasted. One solution to this problem is the use of logarithmic
bins.  This, however, maps the compact region~$[0,1]\times[0,1]$
to~$[-\infty,0]\times[-\infty,0]$, which is inconvenient because of
the missing lower bounds.\par
The more appropriate solution is to use two maps
\begin{equation}
  \begin{aligned}
    \phi : [0,1] &\to     [0,1] \\
           x     &\mapsto y = x^{1/\eta}
  \end{aligned}
\end{equation}
where~$x=x_\gamma$ or~$x=1-x_{e^\pm}$, and to bin the result
equidistantly.  If~$\eta$ is chosen properly (cf.~(\ref{eq:mapping})),
the bin contents will then fall off at the singularity.  The fits in
tables~\ref{tab:param}, \ref{tab:param/TeV}, and~\ref{tab:param/Tesla}
have been performed with~$\eta=5$ and the resulting bin contents can
be read off from
figures~\ref{fig:fit/S-Band/500GeV}--\ref{fig:fit/Tesla/1TeV}.\par 
Using this procedure for binning the results of the simulations, the
popular fitting package \texttt{MINUIT}~\cite{James/Roos:1989:Minuit}
converges quickly in all cases considered. The resulting parameters
are given in tables~\ref{tab:param}, \ref{tab:param/TeV},
and~\ref{tab:param/Tesla}.  Plots of the corresponding distributions
have been shown in figures~\ref{fig:dist}
and~\ref{fig:dist/Tesla}.  It is obvious that an
\textit{ansatz} like~(\ref{eq:beta}) is able to distinguish among the
accelerator designs.  Thus it can provide a solid basis for physics
studies.\par
In figures~\ref{fig:fit/S-Band/500GeV}--\ref{fig:fit/Tesla/1TeV} I
give a graphical impression of the quality of the fit, which appears
to be as good as one could reasonably expect for a simple
\textit{ansatz} like~(\ref{eq:beta}).  Note that the histograms have
non-equidistant bins and that the resulting Jacobians have not been
removed.  Therefore the bin contents falls off at the singularities,
as discussed above.\par
The errors used for the least-square fit had to be taken from a Monte
Carlo~(MC) study.  \texttt{Guinea-Pig} only provides the~$\sqrt n$ from
Poissonian statistics for each bin, but the error accumulation during
tracking the particles through phase space is not available.  The
MC studies shows that the latter error dominates the former, but
appears to be reasonably Gaussian.  A complete MC study of all
parameter sets is computationally expensive (more than a week of 
processor time on a fast SGI).  From an exemplary MC study of a few
parameter sets, it appears that the errors can be described reasonably
well by rescaling the Poissonian error in each bin with appropriate
factors for electrons/positrons and photons and for continuum and
delta.  This procedure has been adopted.\par
The~$\chi^2/\text{d.o.f.}$'s of the fits are less
than~$\mathcal{O}(10)$.
The simple \emph{ansatz}~(\ref{eq:beta}) is therefore very
satisfactory.  In fact, trying to improve the ad-hoc factorized Beta
distributions by the better motivated approximations 
from~\cite{Chen:1992:Beamstrahlung}
or~\cite{Anlauf:1996:Beamstrahlung}, it turns 
out~\cite{Anlauf:1996:Chen_no_good} that~(\ref{eq:beta}) provides a
significantly better fit of the results of the simulations.
The price to pay is that the parameters in~(\ref{eq:beta}) have no
direct physical interpretation.
\nwenddocs{}%
\nwbegindocs{70}%
%%%%%%%%%%%%%%%%%%%%%%%%%%%%%%%%%%%%%%%%%%%%%%%%%%%%%%%%%%%%%%%%%%%%%%
\subsubsection{Generators}
For this version of the parameterizations we need a fast generator of
Beta distributions:
\begin{equation}
  \beta^{a,b}(x)
    \propto x^{a-1}(1-x)^{b-1}
\end{equation}
This problem has been studied extensively and we can use a published
algorithm~\cite{Atkinson/Whittaker:1979:beta_distribution} that is 
guaranteed to be very fast for all~$a$, $b$ such that~$0<a\le1\le b$,
which turns out to be always the case (cf.~tables~\ref{tab:param},
\ref{tab:param/TeV}, and~\ref{tab:param/Tesla}).
\nwenddocs{}%
\nwbegindocs{71}%
%%%%%%%%%%%%%%%%%%%%%%%%%%%%%%%%%%%%%%%%%%%%%%%%%%%%%%%%%%%%%%%%%%%%%%
\subsection{Future Versions}
There are two ways in which the parameterizations can be improved:
\begin{description}
  \item[more complicated functions:] the factorized fits can only be
    improved marginally by adding more positive semi-definite factors
    to~(\ref{eq:beta}).  More improvement is possible by using sums of
    functions, but in this case, the best fits violate the positivity
    requirement and have to be discarded.
  \item[correlations:] the parameterization in
    section~\ref{sec:factorized-beta} is factorized.  While this is a
    good approximation, the simulations nevertheless show correlations
    among~$x_1$ and~$x_2$.  These correlations can be included in a
    future version.
  \item[interpolation:] the parameterization in
    section~\ref{sec:factorized-beta} is based on fitting the
    simulation results by simple functions.  Again, this appears to be
    a good approximation.  But such fits can not uncover any fine
    structure of the distributions.  Therefore it will be worthwhile
    to study interpolations of the simulation results in the future.
    A proper interpolation of results with statistical errors is
    however far from trivial: straightforward polynomial or spline
    interpolations will be oscillatory and violate the positivity
    requirement.  Smoothing algorithms have to be investigated in
    depth before such a parameterization can be released.
  \item[other simulations:]
    besides~\cite{Schulte:1996:Thesis},
    other simulation codes are invited to contribute their results
    for inclusion in the \Kirke/ library.
\end{description}
\nwenddocs{}%
\nwbegindocs{72}%
%%%%%%%%%%%%%%%%%%%%%%%%%%%%%%%%%%%%%%%%%%%%%%%%%%%%%%%%%%%%%%%%%%%%%%
% Local Variables:
% mode:noweb
% noweb-doc-mode:latex-mode
% noweb-code-mode:fortran-mode
% indent-tabs-mode:nil
% page-delimiter:"^@ %%%.*\n"
% End:
\nwenddocs{}%
\nwbegindocs{73}%
%%%%%%%%%%%%%%%%%%%%%%%%%%%%%%%%%%%%%%%%%%%%%%%%%%%%%%%%%%%%%%%%%%%%%%
\section{Conclusions}
\label{sec:conclusions}
I have presented a library of simple parameterizations of realistic
$e^\pm$- and $\gamma$-beam spectra at future linear
$e^+e^-$-colliders.  The library can be used for integration and event
generation.  Emphasis is put on simplicity and reproducibility of the
parameterizations for supporting reproducible physics simulations.
%%%%%%%%%%%%%%%%%%%%%%%%%%%%%%%%%%%%%%%%%%%%%%%%%%%%%%%%%%%%%%%%%%%%%%%%
\subsection*{Acknowledgements}
Daniel Schulte made his simulation code \texttt{Guinea-Pig}
available and answered questions.  Harald Anlauf and Torbjorn
Sj\"ostand have contributed useful suggestions.  The Tesla group at
DESY/Zeuthen made error estimates feasible by donating time on the
multi-headed number cruncher \Hydra/.  The 1996 ECFA/Desy Linear
Collider Workshop got me started and provided support.  Thanks to all
of them.
%%%%%%%%%%%%%%%%%%%%%%%%%%%%%%%%%%%%%%%%%%%%%%%%%%%%%%%%%%%%%%%%%%%%%%%%
%%% \bibliography{circe}

%%%%%%%%%%%%%%%%%%%%%%%%%%%%%%%%%%%%%%%%%%%%%%%%%%%%%%%%%%%%%%%%%%%%%%%%
\appendix
\section{Literate Programming}
\label{sec:litprog}
%%%%%%%%%%%%%%%%%%%%%%%%%%%%%%%%%%%%%%%%%%%%%%%%%%%%%%%%%%%%%%%%%%%%%%%%
\subsection{Paradigm}
\label{sec:paradigm}
I have presented the sample code in this paper using the
\emph{literate programming} paradigm.  This paradigm has been
introduced by Donald Knuth~\cite{Knuth:1991:literate_programming} and
his programs \TeX~\cite{Knuth:1986:TeX} and \MF~\cite{Knuth:1986:MF}
provide excellent examples of the virtues of literate programming.
Knuth summarized his intention as
follows~(\cite{Knuth:1991:literate_programming}, p.~99)
\begin{quote}
  ``Let us change our traditional attitude to the construction of
  programs. Instead of imagining that our main task is to instruct a
  \emph{computer} what to do, let us concentrate rather on explaining
  to \emph{human beings} what we want a computer to do.''
\end{quote}
Usually, literate programming uses two utility programs to produce two
kinds of files from the source
\begin{description}
  \item[\texttt{tangle}] produces the computer program that is
    acceptable to an ``illiterate'' (Fortran, C, etc.) compiler.  This
    process consists of stripping documentation and reordering code.
    Therefore it frees the author from having to present the code in
    the particular order enforced by a compiler for purely technical
    reasons.  Instead, the author can present the code in the order
    that is most comprehensible.
  \item[\texttt{weave}] produces a documents that describes the
    program.  Extensive cross referencing of the code sections is
    usually provided, which has been suppressed in this paper. If a
    powerful typesetting system (such a \TeX) is used, the document
    can present the algorithms in clear mathematical notation
    alongside the code.  These features improve readability and
    maintainability of scientific code immensely.
\end{description}
%%%%%%%%%%%%%%%%%%%%%%%%%%%%%%%%%%%%%%%%%%%%%%%%%%%%%%%%%%%%%%%%%%%%%%%%
\subsection{Practice}
\label{sec:practice}
\Kirke/ uses the \texttt{noweb}~\cite{Ramsey:1994:noweb} system.  This
system has the advantage to work with any traditional
programming language and support the essential features described in
section~\ref{sec:paradigm} with minimal effort.  \texttt{noweb}'s
\texttt{tangle} program only reorders the code sections, but does not
reformat them.  Therefore its output can be used just like any other
``illiterate'' program.\par
The examples above should be almost self-explaining, but in order to
avoid any ambiguities, I give another example:
\nwenddocs{}%
\nwbegincode{74}\moddef{Literate programming example}%
\endmoddef
  \LA{}Code that has to be at the top~{\nwtagstyle{}}\RA{}
  \LA{}Other code~{\nwtagstyle{}}\RA{}
\nwendcode{}%
\nwbegindocs{75}%
I can start the presentation with the first line of the ``other
code'':
\nwenddocs{}%
\nwbegincode{76}\moddef{Other code}%
\endmoddef
line 1 of the other code
\nwendcode{}%
\nwbegindocs{77}%
If appropriate, the first line of the code that has to appear
\emph{before} the other code can be presented later:
\nwenddocs{}%
\nwbegincode{78}\moddef{Code that has to be at the top}%
\endmoddef
line 1 of the code at the top
\nwendcode{}%
\nwbegindocs{79}%
Now I can augment the sections:
\nwenddocs{}%
\nwbegincode{80}\moddef{Other code}%
\plusendmoddef
line 2 of the other code
\nwendcode{}%
\nwbegindocs{81}%
\nwdocspar
\nwenddocs{}%
\nwbegincode{82}\moddef{Code that has to be at the top}%
\plusendmoddef
line 2 of the code at the top
\nwendcode{}%
\nwbegindocs{83}%
The complete ``program'' will be presented to the compiler as
\begin{verbatim}
line 1 of the code at the top
line 2 of the code at the top
line 1 of the other code
line 2 of the other code
\end{verbatim}
The examples in section~\ref{sec:sample-int} show that this
reordering is particularly useful for declaring variables when they
are first used (rather than at the beginning) and for zooming in on
code inside of loops.
\nwenddocs{}%
\nwbegindocs{84}%
%%%%%%%%%%%%%%%%%%%%%%%%%%%%%%%%%%%%%%%%%%%%%%%%%%%%%%%%%%%%%%%%%%%%%%
\section{Fortran Name Space}
\label{sec:name-space}
In addition to the ten procedures and one \texttt{common} block
discussed in section~\ref{sec:usage}
\begin{itemize}
  \item \texttt{circe}, \texttt{circee}, \texttt{circeg}, \texttt{circgg},
  \item \texttt{girce}, \texttt{gircee}, \texttt{girceg}, \texttt{gircgg},
  \item \texttt{circes}, \texttt{circel}, \texttt{/circom/},
\end{itemize}
there are two more globally visible functions which are used internally:
\begin{itemize}
  \item \texttt{circem}: error message handler,
  \item \texttt{girceb}: efficient Beta distribution generator.
\end{itemize}
Even if the \texttt{/circom/} is globally visible, application
programs \emph{must not} manipulate it directly.  The \texttt{circes},
subroutine is provided for this purpose and updates some
internal parameters as well.\par
With features from the current Fortran standard (Fortran90), I could
have kept the last two functions and the \texttt{common} block
private. But since Fortran90 has only
been adopted by a small fraction of the high energy physics community,
I have decided to remain in the confines of Fortran77 (except for the
ubiquitous \texttt{implicit none}).\par
Application programs wishing to remain compatible with future versions
of \Kirke/ must not use \texttt{common} blocks or procedures starting
with \texttt{circe} or \texttt{girce}.
\nwenddocs{}%

\nwbegindocs{85}%
%%%%%%%%%%%%%%%%%%%%%%%%%%%%%%%%%%%%%%%%%%%%%%%%%%%%%%%%%%%%%%%%%%%%%%
\section{Updates}
\label{sec:updates}
Information about updates can be obtained
\begin{itemize}
  \item on the World Wide Web:
    \begin{itemize}
      \item[]\verb+http://crunch.ikp.physik.th-darmstadt.de/nlc/beam.html+
    \end{itemize}
  \item by internet FTP:
    \begin{itemize}
      \item[] host: \verb+crunch.ikp.physik.th-darmstadt.de+
      \item[] user: \verb+anonymous+
      \item[] password: your email address
      \item[] directory: \verb+pub/ohl/circe+
    \end{itemize}
  \item from mailing lists:
    \begin{itemize}
      \item[] \verb+circe-announce@crunch.ikp.physik.th-darmstadt.de+
      \item[] \verb+circe-bugs@crunch.ikp.physik.th-darmstadt.de+
      \item[] \verb+circe-discuss@crunch.ikp.physik.th-darmstadt.de+
    \end{itemize}
    Subscriptions are available from
    \begin{itemize}
      \item[] \verb+majordomo@crunch.ikp.physik.th-darmstadt.de+
    \end{itemize}
\end{itemize}
Contributions of results from other simulation programs and updated
accelerator designs are welcome at
\begin{itemize}
  \item[] \verb+Thorsten.Ohl@Physik.TH-Darmstadt.de+
\end{itemize}
%%%%%%%%%%%%%%%%%%%%%%%%%%%%%%%%%%%%%%%%%%%%%%%%%%%%%%%%%%%%%%%%%%%%%%%%
\end{document}